\documentclass[sigplan]{acmart}

\usepackage{scalerel}
\newcommand\vfrac[2]{\ThisStyle{%
  \setbox0=\hbox{$\SavedStyle#1#2$}%
  \setbox2=\hbox{$\SavedStyle X$}%
  \ifdim\ht0>\ht2\setlength{\ht0}{\ht2}\fi%
  #1\mathord{\stretchto{\raisebox{2.3\LMpt}{$\SavedStyle/$}}{\ht0}}#2}}

\usepackage{breqn}
\usepackage{caption}
\usepackage{subcaption}
\usepackage{url}

\usepackage{mathrsfs}
\usepackage{amsmath}
\usepackage{booktabs} 
\usepackage{setspace}
\usepackage{setspace}
\usepackage{subcaption}
\usepackage{verbatim}
\usepackage{courier}
\usepackage{amsmath}
\usepackage{listings}
\usepackage{enumitem}
\usepackage{xspace}
\usepackage{enumitem}

\usepackage[colorinlistoftodos]{todonotes}
\usepackage{xcolor}
\usepackage[most]{tcolorbox}
\usepackage{todonotes}

\usepackage{xr}
\definecolor{mygreen}{rgb}{0,0.6,0}
\usepackage{multirow}
\usepackage{float}
\usepackage{graphicx}
\usepackage{caption}
\usepackage{subcaption}

\newcommand{\name}{\texttt{PARIS}\xspace}
\newcommand{\subheader}[1]{ \textbf{#1}}

\newcommand{\frameworkname}{Estimator}
\setcopyright{none}
\renewcommand\footnotetextcopyrightpermission[1]{} 
\begin{document}
\title{PARIS: Predicting Application Resilience Using Machine Learning
}

\author{Luanzheng Guo}
\affiliation{
  \department{EECS}              
  \institution{UC Merced}            
}
\email{lguo4@ucmerced.edu}          

\author{Dong Li}
\affiliation{
  \department{EECS}             
  \institution{UC Merced}           
}
\email{dli35@ucmerced.edu}         

\author{Ignacio Laguna}
\affiliation{
  \department{EECS}             
  \institution{LLNL}           
}
\email{ilaguna@llnl.gov}         

\begin{abstract}
Extreme-scale scientific applications can be more vulnerable to soft errors
(transient faults) as high-performance computing systems 
increase in scale. The common practice to evaluate the resilience to faults
of an application is random fault injection, a method that can be highly time
consuming. While resilience prediction modeling has been recently proposed
to predict application resilience in a faster way than fault injection,
it can only predict a single class of fault manifestation (SDC) and there is no evidence demonstrating
that it can work on previously unseen programs, which greatly limits its re-usability.

We present \name, a resilience prediction method that addresses the problems 
of existing prediction methods using machine learning. Using
carefully-selected features and a machine learning model, our method is able
to make resilience predictions of three classes of fault manifestations (success, SDC, and interruption)
as opposed to one class like in current resilience prediction modeling. The generality of our approach allows us to make prediction on new applications, i.e., previously unseen
applications, providing large applicability to our model.
Our evaluation on 125 programs shows that \name 
provides high prediction accuracy, 82\% and 77\% on average for 
predicting the rate of success and interruption, respectively, while the state-of-the-art resilience prediction model cannot predict them. 
When predicting the rate of SDC, \name provides much better accuracy than the state-of-the-art (38\% vs. -273\%).  
\name is much faster (up to 450x speedup) than the traditional method (random fault injection). 





\end{abstract}


%
%
\maketitle

\section{Introduction}
\label{sec:intro}

High-performance computing (HPC) systems are increasingly used 
to run large scientific applications that simulate real-world phenomena.
These applications are expected to compute precise and correct 
numerical answers for a large set of science and engineering problems. 
As HPC systems increase in scale, they become 
more susceptible to soft errors~\cite{baumann2005radiation} 
(also known as transient faults) due to feature size shrinking, 
lower voltages, and increasing densities in hardware 
infrastructures~\cite{asplos15:vilas}. As a result, scientific
applications running at extreme scale must apply different
resilience methods to tolerate frequent soft errors.
Applying these methods to a given application often requires 
a deep understanding of the degree of resilience 
of the application.

The common practice to study application resilience
to errors in HPC system is \textit{ 
Fault Injection (FI)}~\cite{dsn12:xu,dsn14:wei,sc14:cher,
europar14:calhoun,micro16:hari,
sc17:Georgakoudis,ispass17:hari,isca17:Kaliorakis,bifit:SC12}.
This approach uses a large amount of random FIs,
each of which randomly selects an instruction, and then 
triggers bit flips at the instruction input or output operands 
during application execution. Statistical results are then 
used to quantify application resilience. A typical 
analysis is, for example, to measure the percentage 
of times a fault yields \textit{silent data corruption} 
(SDC) in the output of the program.

While FI works in practice and is widely used
in resilience studies, a key problem of this approach is 
that it is highly time consuming, and as a result it 
is usually applied to limited scenarios, for example, 
on applications that run for a short period of time and/or 
single-threaded codes. To illustrate the problem, consider 
an application that runs for 6 hours---a typical
execution time for a large-scale scientific simulation. Using
statistical analysis (e.g., using~\cite{leveugle2009}), the number of 
random FIs to obtain a low margin of error 
(e.g., 1\%-3\%) is in the order of thousands of injections. 
Thus, the total FI campaign could last 
several days. For multi-threaded or multi-process 
applications, this time is much higher since random 
faults must be injected in different threads or processes.

Since FI is too time consuming to measure
application resilience, recent work proposed the idea of
\textit{predicting} application resilience using an error-propagation
model, as opposed to measuring application resilience via
FI. The idea is to build a model that explains
how errors in instructions propagate to the program output and then use
the model to estimate the percentage of times SDC occurs.
Here, resilience prediction avoids running the code multiple 
times (as in traditional FI) at the price of 
being sometimes less accurate than FI.
The most recent work on this domain is
\texttt{Trident}~\cite{dsn18:li}, which uses an 
analytical model to predict the 
rate of SDC for a given application and input.

While the \texttt{Trident} approach is a useful first 
step in the direction of application resilience prediction, it
has some limitations. First, the approach predicts only a specific
class of fault manifestation, SDC.
Usually when analyzing the resilience of an application,
scientists are interested in understanding at least three
classes of fault manifestations: 
(1) SDC, (2) interruptions (i.e., crashes or hangs), and (3) success
(i.e., the fault was benign and did not affect the program output).
Because of the low-level analytical modeling approach
of Trident, it cannot predict all of these three cases.
Second, resilience predictions are made on the
program that was used to build the model and there is no evidence
demonstrating
that it can work on previously unseen programs, which greatly limits the re-usability of the model.

\subheader{Paper Contributions.}
We present \name, an approach for fast and accurate prediction 
of application resilience. \name avoids the time-consuming process
of randomly selecting and executing many injections
that FI incurs. Using machine learning and a large 
set of representative programs and kernels as the input, we build a generic 
model that explains how error masking
and error propagation occur within code regions. Once the model is trained, it can be used to 
predict three classes of fault manifestations---SDC, interruptions 
and success---for a new, previously unseen application, 
addressing the major limitations of \texttt{Trident}.

Our machine learning modeling is unique and is based 
on several key principles.
First, we aim at building a model that captures the implicit relationship between 
application characteristics and application resilience, which 
is difficult to capture for analytical models, such as~\cite{dsn18:li}.
Second, we carefully choose application characteristics 
as features for the model; we select features that are 
directly related to application resilience and that 
capture the order of execution of instructions as we found that the
latter is critical in accurately modeling error propagation.
Third, we perform sophisticated 
feature reduction to enable efficient model training.
Fourth, we perform a large model-selection search among
18 different widely-used models---while regression
is the most natural choice of modeling for our problem,
there are a number of regression models, thus we must 
answer which model can provide the best accuracy.

In summary, our contributions are as follows.

\begin{itemize}[leftmargin=*]
\item We present \name, the first machine learning-based approach
to predict application resilience; we find that our
approach is up to 450x faster than the traditional 
FI approach.

\item We describe a framework to build machine leaning models
that can predict more classes of fault manifestations than the
state-of-the-art resilience prediction method, 
\texttt{Trident} \\
~\cite{dsn18:li}: three in our case 
(SDC, interruptions, and success) versus one (SDC) 
on \texttt{Trident}.
Our prediction accuracy for 
SDC 
is better than \texttt{Trident} (38\% vs. -273\% on average).
Our prediction accuracy for success and interruptions prediction
is 82\% and 77\%, with a variation of 0.02 and 0.05, respectively,
while \texttt{Trident} cannot predict those cases.

\item We design and demonstrate that our modeling method can predict
application resilience on previously unseen applications, i.e.,
applications that were not used on the modeling phase, which current
methods cannot do.
\end{itemize}

Our evaluation methodology is solid and comprehensive. We use in total
of 125 programs, 18 machine learning models and perform 375,000 
FIs to compare our results to those of 
traditional random FI, using a confidence interval 
of 99\% and a margin error of 1\%. To the best of our knowledge, we are the first on 
providing such a compressive evaluation for a resilience 
prediction method.

\section{Background} 

We present useful background information and definitions
before presenting our approach.

\subsection{Fault Model}
\label{fault_model}

We consider soft errors~\cite{baumann2005radiation}, 
i.e., transient faults that escape
from hardware protection and propagate to the application level.
These errors manifest as bit flips in registers and memory locations. 
Corrupted registers and memory cells are consumed 
by the application. We refer to any of these registers and memory 
locations as a \textit{location} in the paper. 

We focus on single-bit errors, not multi-bit errors.
The reason behind this is that
(1) single-bit errors are the most common soft errors---multi-bit 
errors rarely happen in large-scale systems~\cite{sc16:bautista};
(2) in many cases, multi-bit errors have the similar impact on the 
application as single-bit errors~\cite{dsn17:Sangch}. 

\subheader{Fault injection.}
We use PINFI~\cite{dsn14:wei} to perform fault 
injections into programs. Comparing with LLFI~\cite{selse2013llfi} and REFINE~\cite{sc17:Georgakoudis} (two common FI tools), PINFI~\cite{dsn14:wei} is more accurate 
than LLFI and comparable to 
REFINE for FI. 

\begin{figure*}[th!]
	\begin{center}
		\includegraphics[width=0.97\textwidth]{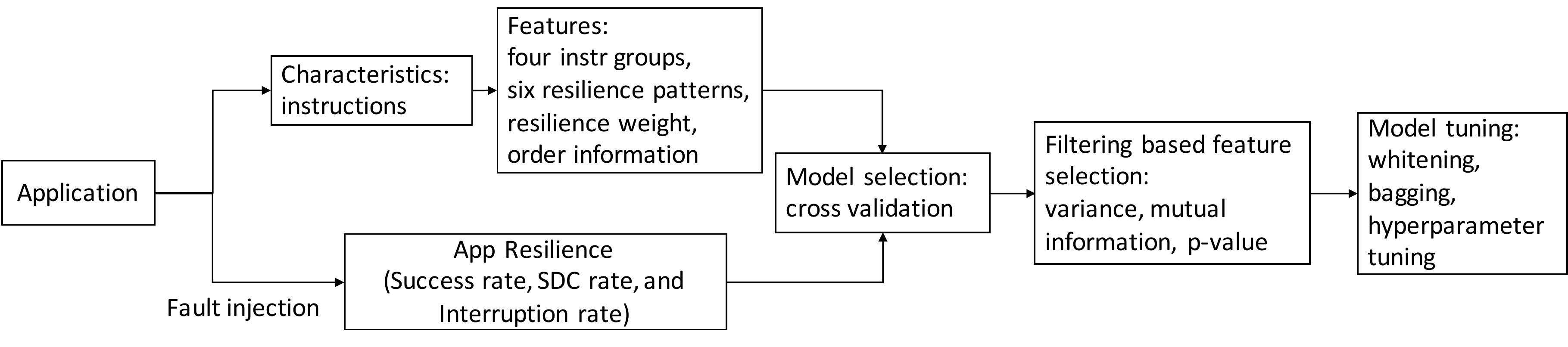} 
		\caption{Overview of \name and the workflow of the training process in our machine learning model.}
		\label{fig:flow_chart}
	\end{center}
\end{figure*}

\subsection{Application Resilience}
\label{three_metrics}

We run FI campaigns to measure the application 
resilience. A FI campaign contains 
many FIs. In each FI, 
a single-bit error is injected into an input/output 
operand of an instruction. A \textit{location} 
can be an operand of any instruction. 

We classify the outcome, or \textit{manifestation}, 
of programs corrupted by bit flips into three 
classes: success, SDC, and interruption:
\begin{itemize}[leftmargin=*]
\item \textbf{Success}: the execution outcome is exactly the same 
as the outcome of the fault-free run. The execution 
outcome can also be different from the outcome of the fault-free run, 
but the execution successfully passes the result verification 
phase of the application.

\item \textbf{SDC}: the program outcome
is different from the outcome of the fault-free execution,
and the execution does not pass the result verification 
phase of the application.  

\item \textbf{Interruption}: the execution does not 
reach the end of execution, i.e., it is interrupted 
in the middle of the execution, because of an 
exception, crash, or hang.

\end{itemize} 

\subheader{Rates.}
To quantify the application resilience in a given FI campaign, we measure the rate of each of the three
classes of manifestations described above.
In particular, we use the formula:
\begin{equation}
\label{eq:rates}
\text{\#Manifestations}/N,
\end{equation}
where \text{\#Manifestations} is the number of times
a given class of manifestation (success, SDC or interruption)
occurs, and $N$ is the number of FIs 
performed in a FI campaign. In this paper, 
we consider the rates of success, SDC and interruption as
metrics to quantify application resilience.

\subsection{Features and Machine Learning Models}
\subheader{Features.}
A feature is an application characteristic. 
Multiple features construct a \textit{feature vector}, $v$,
which is used as the input for our machine learning models. 
Choosing discriminative features is an essential 
task for building an effective machine learning model. 
However, irrelevant and redundant features can affect 
the modeling accuracy, thus we must perform feature selection
on the modeling process.

\subheader{Models.}
There are two classes of machine learning models: 
classification and regression. Our problem is naturally 
a regression problem since the manifestation rates 
are real numbers.
More formally, our problem is to find a model $f()$, for which given an feature vector $v$ corresponding to an application $A$, $f(v)$ gives us the rates of SDC, interruption, and success for $A$. Note that the SDC, interruption, and success rates are
real numbers between 0.0 and 1.0. Since they are mutually
exclusive events, the addition of them for
a given application is 1.0.

\subheader{Prediction Accuracy.}
After our model is trained, it is used to predict application resilience.
To estimate the prediction accuracy of the model,
we compute the relative error between the predicted application 
resilience and the application resilience that we observe 
by performing FI. The prediction accuracy 
$P_{\textit{accuracy}}$ is defined as follows:

\begin{equation}
\label{eq:3}
P_{\textit{accuracy}}  = 1 - \frac{|P_{\textit{rate}}-O_{\textit{rate}}|}{O_{\textit{rate}}},
\end{equation}
where $P_{\textit{rate}}$ is the predicted rate (of either the
success, SDC or interruption), and $O_{\textit{rate}}$ is the observed 
rate by performing FI.
A perfect model has a $P_{\textit{accuracy}}$ of 1.0.

\textbf{Training and Testing Phases. }
The modeling process of machine learning includes training and testing. 
We use a set of representative applications 
to train the model---once it is trained, the model is used
to predict, or test, the manifestation rates on new applications. 
We call the applications used for training and testing 
the training dataset and the testing dataset, respectively.

\section{Overview}
\label{sec:overview}

We give a high-level overview of \name. Figure~\ref{fig:flow_chart} depicts the workflow of 
the training process in \name.
The most challenging part 
of the training process
is to construct features relevant to application resilience and producing high modeling accuracy. 

\textbf{Features Construction.}
We use instruction type and number of instruction instances for each type as features. 
A static instruction in a program
has an instruction type (opcode), and 
can be executed many times, each of which is an \textit{instruction instance}. 
Using the number of instruction instances for each instruction type as a feature will result in too many features, which demands a large training dataset. To reduce the number of features, we group all instruction types (65 in total) into four groups: control flow instructions, floating point instructions, integer instructions, and memory-related instructions. 
For each instruction group, we count the number of instruction instances as a feature. 

Furthermore, we use six resilience computation 
patterns proposed in~\cite{fliptracker17} as features. 
Among the six patterns, four of them (conditional statement, shifting, data truncation, and data overwriting) are 
individual instructions that cannot be grouped into the four instruction groups. Two of them (dead locations and repeated addition) include multiple instructions and those instructions together (not individual instructions) contribute to application resilience.

Counting dead 
locations and repeated addition from the dynamic instruction trace is challenging because we must repeatedly search within the trace to find correlation between instructions. 
To detect dead locations, we use a technique that caches intermediate results of trace analysis to avoid repeated trace scanning. To detect repeated addition, we build a data dependency graph for addition instructions. Such graph enables easy detection of repeated addition.

Because different instruction instances can have 
different capabilities to tolerate errors, even though 
they have the same instruction type (or the same resilience computation pattern), we introduce
\textit{resilience weight} when counting instruction instances. 
The resilience weight gives each instruction instance
a weight quantifying the possible number of single-bit errors
that can be tolerated by the instance.

Furthermore, we introduce execution order information as a feature. 
We notice that instruction order
can affect the application 
resilience. 
However, representing the order information of all instruction instances as a feature is a challenge.
We use N-gram~\cite{brown1992cl,cavnar1994n}, a technique commonly used for processing speech data, to capture the order information. 


\textbf{Machine Learning Techniques. }
There are many machine learning models we can use for \name.
We use a model selection
method~\cite{kohavi1995study} to find
the machine learning model with the highest prediction accuracy.
We also use the filtering-based feature
selection~\cite{das2001filters} to filter out 
irrelevant and redundant features.
As used in the existing machine learning 
work, we use model tuning 
techniques including whitening, bagging, 
and hyperparameter tuning to improve the prediction accuracy of our model.

\textbf{Use of \name. }
To use \name, users do not need doing FI. 
Instead, users generate a dynamic instruction trace and feeds it to \name.
\name will output a numerical value, which is the predicted success, SDC, or interruption rate.

\section{Design}
We describe the design details in this section.

%
\subsection{Feature Construction}
\label{feature_construction}

To construct features for the regression models, we have the following requirements: 
(1) The features should be relevant to application resilience. (2) The number of features should be small enough. Ideally, the number should be much smaller than the number of applications for training to avoid under-determination of the model.
(3) We should avoid redundant and
irrelevant features. 
Those features can lower prediction accuracy.
We describe our feature construction following the above requirements in this section. 

\subsubsection{Instruction Groups}

The first features we introduce into the models are the instruction type and number of instruction instances in each type. 
These features are highly relevant to application resilience. 
For example, recent studies~\cite{elliott2013quantifying,menon2018d,dsn18:li}
reveal that floating point instructions 
make application resilient to faults, because the faults in mantissa bits of floating point numbers are often ignorable by the application (especially HPC applications).
Load/store instructions also has significant impact on application resilience, because the computation following load/store instructions may need those loaded/stored values.

We use the LLVM compiler~\cite{lattner2004llvm} to 
build applications, which is architecture  independent, allowing us to build a more general and reusable model for resilience prediction.
We enumerate all LLVM IR instructions and get 65 instruction types. 
However, building a feature vector with 65 instruction types 
will lead to at least 65 unknown 
variables in the solution space.
To fill up the solution space, it could easily require thousands of applications~\cite{chen2017nips,sigkdd2000Domingos} to train the models. This violates the requirement (2) and can be time-consuming to find the optimal solution.

To address this problem, we group 65 instruction types into four groups 
 based on the functionality of instructions to reduce the number of features. 
Instruction functionality relies on
instruction types; different instruction types have different
impacts on application resilience. 
For example, we group control flow related instructions (e.g., \texttt{Br} and \texttt{Select}) into a group,
and group floating point instructions (e.g., \texttt{Fadd} and \texttt{Sine}) into a group. 
Table~\ref{tab:grouping} lists the four groups, including control flow instructions, floating point instructions, integer instructions, and memory-related instructions.

For each instruction group, we count the number of instruction instances from the dynamic instruction trace, and then normalize the number based on the total number of instruction instances. 
We normalize the number as a feature to make the feature value independent of the size of the dynamic instruction trace.

\begin{table}
\begin{center}
\caption {Four groups of instruction types and three resilience patterns as features to build the machine learning models.}
\label{tab:grouping}
\tiny
\begin{tabular}{|p{2.4cm}|p{5.4cm}|}
\hline
\textbf{Group Name} 	& \textbf{Instruction types} 	          \\ \hline \hline
Control Flow Instructions (CFI) & Br, Indirectbr, Select, PHI, Fence, DMAFence, Call
\\ \hline
Floating Point Instructions (FPI)   	       & Fadd, Fsub, Fmul, Fdiv, Frem, Cosine, Sine  	\\ \hline
Integer Instructions (II)    & add, sub, mul, Udiv, Sdiv, Urem, Srem    \\ \hline
Memory-related Instructions (MI)            & Load, Store, DMAStore, DMALoad, Getelementptr, ExtractElement, InsertElement, ExtractValue, InsertValue, FPToUI, FPToSI, UIToFP, SIToFP, PtrToInt, IntToPtr, AddrSpaceCast	\\ \hline \hline
\textbf{Pattern name}       &  \textbf{Instruction
types} \\ \hline
Condition             & ICmp, FCmp, Switch, And, Or, Xor  \\ \hline
Shift & Shl, LShl, AShl  \\ \hline
Truncation & Trunc, ZExt, Sext, FPTrunc, BitCast, FPExt   \\ \hline
\end{tabular}
\end{center}
\end{table}

\subsubsection{Using Resilience Computation Patterns as Features}
The recent work~\cite{fliptracker17} finds six resilience computation patterns strongly
related to application resilience.
The resilience computation pattern is defined as 
combinations or sequences of computations that make an application naturally resilient. 
These six patterns are dead locations, repeated addition, conditional statements, shifting, data truncation, and data overwriting. 
We introduce the patterns of dead locations and repeated addition as features, because the two patterns include multiple instructions and those instructions together (not individual instructions) contribute to application resilience; 
The other four patterns are individual instructions and 
could fall into the four instruction groups. However, 
we use them separately as features, because of their significance to application resilience~\cite{fliptracker17}.

We describe how to introduce the six patterns as features as follows. A pattern can be repeatedly executed by the application. We name the execution of a specific pattern within the application as the \textit{pattern instance}.
To introduce these six patterns as features, we could simply count the number of the pattern instances for each pattern. However, doing so has a couple of challenges.

First, counting the number of the pattern instances for dead locations and repeated addition can be time-consuming, because we have to find correlations between instructions to determine if the location is dead or if the addition repeatedly happens to the same variable. 
Doing so requires iteratively scan the dynamic instruction trace. We discuss how to efficiently count the number of the pattern instances for the two patterns in Section~\ref{sec:count_dead} and Section~\ref{sec:count_rep_add}, respectively. 


Second, for the patterns of conditional statements, shifting, data truncation and data overwriting that are represented as individual instructions (see the last three rows in Table~\ref{tab:grouping} for these instructions), simply counting the number of pattern instances cannot discriminate the fault tolerance capabilities of different pattern instances. For example, the fault tolerance capability of the ``shifting'' pattern (a pattern involving a \textit{shift} instruction) depends on how many bits are shifted. A \textit{shift} instruction instance shifting three bits can tolerate three single-bit errors, while a \textit{shift} instruction instance shifting one bit can tolerate one single-bit error. To distinguish the fault tolerance capabilities of different instruction instances, we introduce weights (named \textit{resilience weight}) for counting instances of the patterns.
Besides introducing weights for the patterns of conditional statements, shifting, data truncation and data overwriting, we also introduce weights to instructions whose instances can also have different fault tolerance capabilities. 
We describe how we determine weights for instruction instances in Section~\ref{sec:res_weight}.

\subsubsection{Extracting the Feature of Dead Location}
\label{sec:count_dead}

Dead locations refer to those locations that have short live time. 
While the errors in  those locations can propagate to one (or a few) other locations, many of the dead locations are not used anymore. As a result,
the total number of corrupted locations in the program can decrease, because of those short live locations. 
A code region with a higher percentage of dead locations (i.e., the number of dead locations over the number of total locations) has higher resilience.

To efficiently detect dead locations and calculate the percentage of dead locations, 
we split the dynamic instruction trace into chunks and pre-process the chunks before detecting dead locations.
During the trace pre-processing, we analyze instructions in each chunk and record names of locations within the chunk. This location record is saved in an array for each chunk.
To determine if a location in a chunk
is a dead location, we only need to check whether the same location is used in any following chunks by examining the arrays.  
If the location is not used, then the location is a dead location. In essence, the arrays for chunks save instruction analysis results to avoid repeatedly scanning the trace and analyzing instructions.
For each chunk, we 
normalize the number of dead locations by the total
number of locations within the chunk and
compute the percentage of dead locations for the chunk.
We use the average dead location rate of all chunks as a feature.


\subsubsection{Extracting the Feature of Repeated Addition}
\label{sec:count_rep_add}
\begin{figure}
	\begin{center}
\includegraphics[height=0.15\textheight]{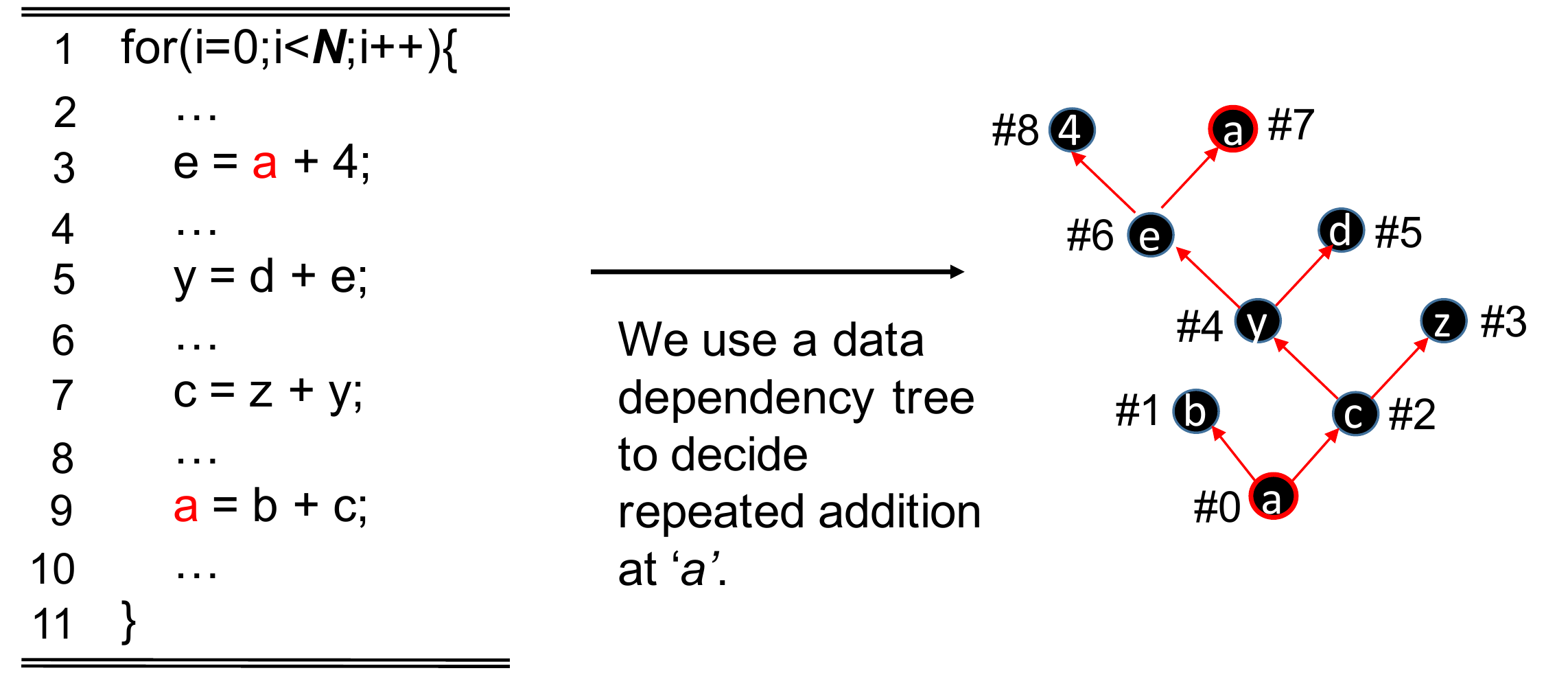} 
		\caption{An example to detect repeated additions. }
		\label{fig:repeated_addition}
	\end{center}
\end{figure}

Repeated addition refers to the addition repeatedly happening to a variable, such that the error in the variable can be amortized. 
To decide if an addition instruction is a part of repeated addition, we must first
decide if the addition instruction is 
involved in a self addition. The self addition is that a location adds other location(s) to itself.
The pseudo code in Figure~\ref{fig:repeated_addition}
gives an example of self addition. 

To detect a self addition, we first construct a data dependency graph for addition operations. 
A node in the graph is a location; edges between graph nodes represent data dependency.
Then, given an addition instruction, we examine the output operand of the addition instruction, and decide if the location (the output operand) is a source operand of a previous addition operation by backward traversing the graph. 

Figure~\ref{fig:repeated_addition} illustrates how a data dependency graph looks like and how a self addition is found. 
In the example, we have four addition statements (operations) in a \textit{for} loop. The location $a$ appears as the output of the last addition statement ($a=b+c$ in Line 9). To determine if the addition statement is involved in a self addition, we find the node 0 corresponding to $a$ in the data dependency graph. We traverse the graph backward, and find $a$ appears in a previous node, the node 7.
The node 7 corresponds to a source operand of a previous addition statement ($e=a+4$). Hence, a self addition is detected. A repeated addition is composed by a number of self additions. 

 
To use repeated addition 
as a feature, we normalize the 
number of repeated addition instances 
by total number of instruction instances. 
This makes the feature value independent of the size of the dynamic instruction trace.

\subsubsection{Resilience Weight}
\label{sec:res_weight}

Given an instruction, all bit locations of its input and output operands are subject to error corruption. The resilience weight ($\mathcal{R}{es}$) of an instruction is defined as follows. 

\begin{equation}
\label{eq:res}
\mathcal{R}{es}  = \frac{\#bit\ locations\ that\ tolerate\ errors}{\# of all\ bit\ locations}
\end{equation}

Using the \textit{right-shift} instruction as an example.
The instruction has three 8-bit operands and in total 24 locations.
Assume that an instance of the instruction shifts four least significant bits of an operand.
The shifted four bits can tolerate four single-bit errors. Also, the eight bits in the output operand of the instruction can tolerate errors because of the result overwriting in the output operand. Hence, in this example,
the resilience weight for this instruction instance 
 is (4+8)/24 = 0.5.

For any floating point and integer instruction, the bit locations that can tolerate errors are the bit locations of the output operands, because we expect errors in the output operands can be overwritten by the instructions. 

When counting the number of instruction instances or the number of pattern instances, we use the weights to account the numbers. 
\noindent \textbf{Putting All Together.}
As a result of the above feature construction, we construct a feature vector of ten features, formulated in Equation~\ref{eq:feature_vector}. The notation of the equation can be found in Table~\ref{tab:grouping}.

\begin{dmath}
\label{eq:feature_vector}
\mathcal{F}^{ave}_{10} = [CFI, FPI, II, MI, \\Condition, Shift, Truncation, DO, DLR, RA]
\end{dmath}

We call $\mathcal{F}^{ave}_{10}$ the \textit{foundation feature vector} and call the ten features \textit{foundation features} in the rest of the paper. 

\subsection{Including Instruction Order}
The foundation features are not good enough to achieve high prediction accuracy. 
In particular, the foundation features lack instruction order (i.e., the execution order) information. Capturing the instruction order is important, because it matters to error propagation. 

\begin{figure}
	\begin{center}
\includegraphics[height=0.14\textheight]{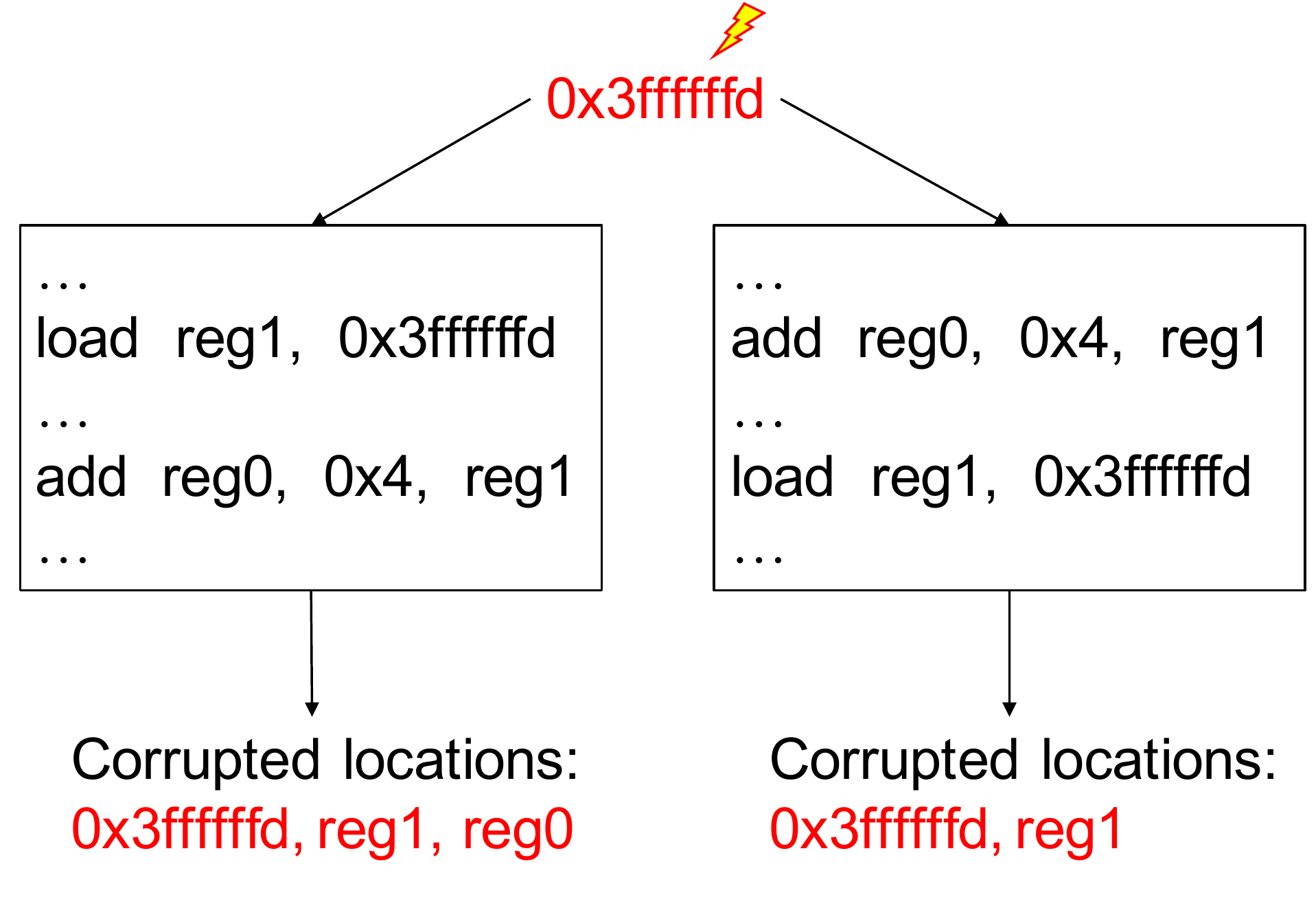} 
		\caption{An example to show that the execution order of instructions matters to error propagation.}
	\label{fig:why_order}
	\end{center}
\end{figure}

To give an intuition of why the execution order matters, we use a simple example shown in Figure~\ref{fig:why_order}. In this example, we have a $load$ instruction and an $addition$ instruction. Assume that an error happens on a memory address 0x3ffffffd. 
If the $load$ instruction happens first, then the erroneous value in the memory address can propagate to the locations $reg1$ and $reg0$. 
but if the $addition$ instruction happens first, then the erroneous value in the memory address can only propagate to the location $reg1$. This  example is a demonstration of how the execution order matters to error propagation. 

To introduce execution order information into the feature vector, we use the ``N-gram'' technique. The N-gram is a technique used in computational linguistics. It can work on a sequence of streaming words, and predict next word using sequences of previous words. N-gram can capture the word order information. In particular, every $n$ continuous words composes a $n-$gram ($n=1, 2, 3,...$). We can introduce the order information into features by using the N-gram. 

In particular, we partition the dynamic instruction trace into chunks (each chunk is a gram). Each chunk is treated as a ``word'', and the sequence of chunks
is processed as the sequence of words. 
For each chunk, we collect ten foundation features, and build a foundation feature vector of size ten for each chunk. Then, we build an average foundation feature vector (denoted as $\mathcal{F}^{ave}_{10}$) whose feature values are the average values of foundation feature vectors of all chunks. 

Furthermore, we combine every two chunks to build a 2-gram (or bigram in the language of N-gram). For each bigram, we combine two foundation feature vectors to build a 2-gram feature vector of size 20. After that, we build an average 2-gram feature vector (denoted as $\mathcal{F}^{ave}_{20}$) for all bigrams.
$\mathcal{F}^{ave}_{20}$ is the average value of all 2-grams feature vectors; $\mathcal{F}^{ave}_{20}$ has a size of 20.

\begin{figure}
	\begin{center}
			\includegraphics[height=0.21\textheight]{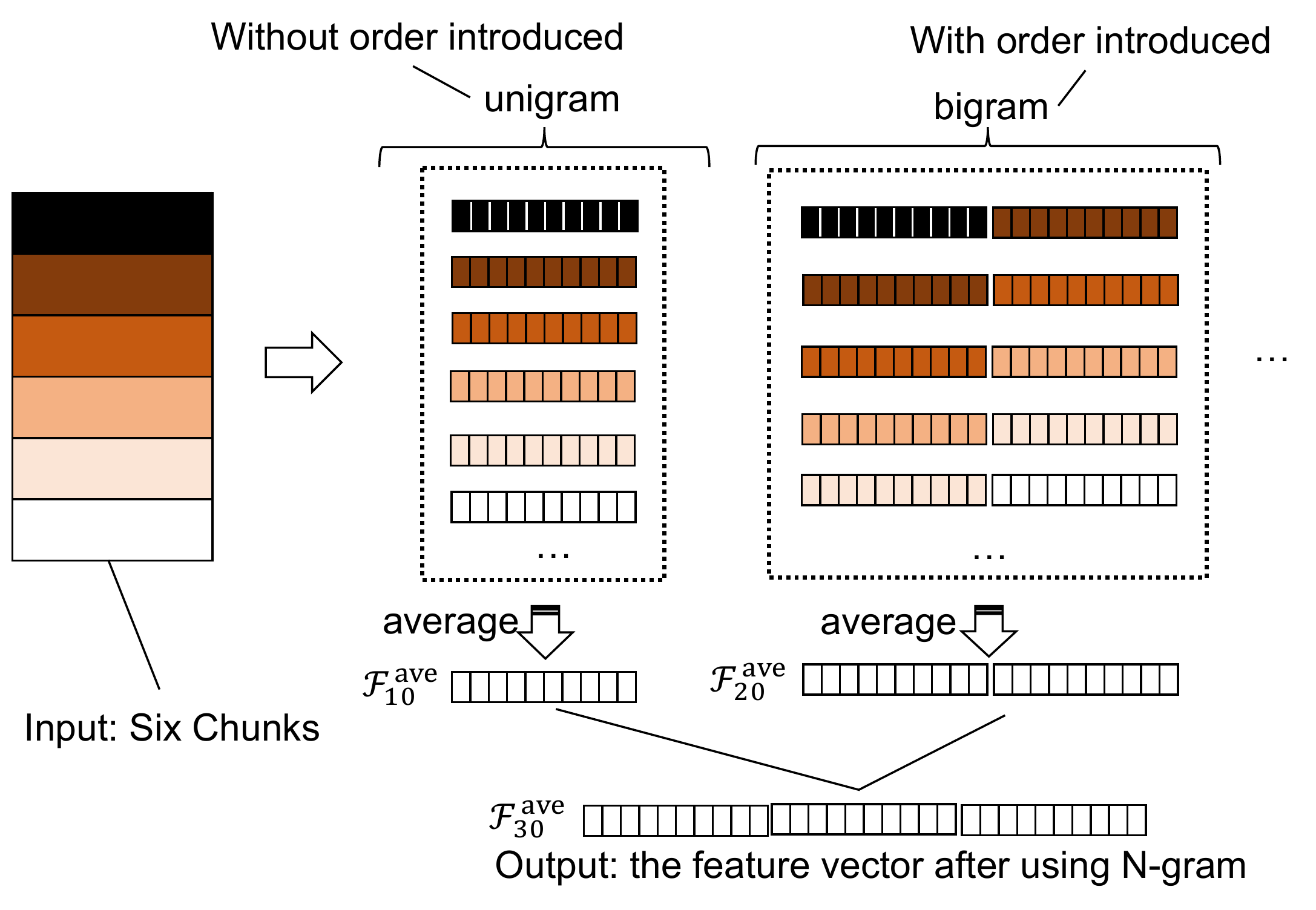}
		\caption{Applying the N-gram technique to introduce the execution order information.}
		\label{fig:n-gram}
	\end{center}
   \vspace{-25pt}
\end{figure}

After that, we have $\mathcal{F}^{ave}_{10}$ of size 10 and $\mathcal{F}^{ave}_{20}$ of size 20. The new feature vector with the execution order information is a combination of $\mathcal{F}^{ave}_{10}$ and $\mathcal{F}^{ave}_{20}$. The new feature vector has a size of 30. We 
denote the new feature vector 
$\mathcal{F}^{ave}_{30}$
Figure~\ref{fig:n-gram} depicts how we build the feature vector with the execution order information included. 

We do not consider trigram (i.e., 3-gram) or higher gram, because 
common practices~\cite{pei2014max,chen2015gated} demonstrate that 
there is no need to use higher grams
than bigram. 
In~\cite{chen2015gated}, bigram achieves better accuracy than trigram.
Using trigram or higher grams does not provide much improvement in prediction accuracy, but dramatically increases the feature vector size and increases the complexity of model training. 

\subsection{Models Selection}
\label{design:model_selection}

There are tens of regression models. Each of them has pros and cons, and can be fit to different scenarios. We explore 18 most common regression models. 

We use cross-validation to evaluate 18 regression models on the training dataset to select the best models.
CV partitions the dataset into $p$ folds. $q$ of $p$ folds are used for training, while the remaining $p-q$ folds are used for testing. There are $p/(p-q)$ rounds of training/testing. In each round, different $p-q$ folds are used for testing. 
We choose the regression models that have the highest prediction accuracy (on average) among all testing dataset. 

In our study, we choose top five regression models based on CV. 
The five models are 
shown in Table~\ref{tab:err_vari}.
Among the remaining 13 models, 12 of them have very poor prediction accuracy (the prediction accuracy is even negative, based on the calculation of Equation~\ref{eq:3}); one of them (i.e., the decision tree) is just a special case of one of the top five (the random forest) in nature. 
For the five selected models, we improve their model accuracy as follows.

\subsection{Feature Selection}
\label{sec:featuer_selection}
After the five models are selected, we explore the possibility to reduce the feature vector size for each of those models. Reducing the feature vector size is useful to eliminate those irrelevant and redundant features to improve modeling accuracy.

We use three common techniques to select features: variance, p-value, and mutual information. 
Simply speaking, the variance of a feature measures the variance of feature values across different input code;
The p-value is metric that that measures the significance level between a feature and the modeling result (i.e., the success, 
SDC, or interruption rate); 
The mutual information measures the mutual dependency between a feature and the modeling result. 

Using the above method, we sort features into a list.
In total, we have three lists, each of which corresponds to one of the three techniques.
We design a voting strategy based on the combination of the three lists. In particular, each feature in a list has an index.
For each feature, we add its three indexes to get a global index.
We sort the features again based on the global index. 
We choose
the best $k$ (where $k=2, 3, ..., 30$) features according to their prediction accuracy.
Such voting strategy and feature selection algorithm~\cite{liu2016tbe,zhang2006bmcb,tsymbal2005if} are common in machine learning.

\subsection{Model Tuning}
After the feature selection process, we further tune the five models.
We choose the one with the highest prediction accuracy as the final model.
We use the following tuning techniques
for model tuning. 

\textbf{Whitening:}
Whitening~\cite{coates2011analysis} is commonly used for avoiding domination effects of any features for better 
generalization to improve the modeling accuracy. 

\textbf{Bagging (Model Averaging):}
Bagging~\cite{domingos2000icml} is often used for reducing the variance in the training data, so that we can eliminate the effect of bad outliers. 

\section{Implementation}
\label{sec:impl}

\textbf{Dataset Construction.}
We have multiple requirements on creating training and testing dataset. (1) The training dataset must be large to avoid model underdetermination (i.e., the evidence available is insufficient to identify which belief one should hold about that evidence.);
(2) applications used to generate training and testing dataset must have diverse computation and have diverse resilience characteristics.
(3) Applications used to generate training dataset must have explicit result verification phases. 
Having those phases allows us to easily determine the fault manifestation (SDC, interruption, and success). 

We use representative benchmark suites and scientific applications to create the testing dataset, including NAS parallel benchmark suite~\cite{nas}, PARSEC benchmark suite~\cite{bienia2008parsec},
CORAL benchmark suite~\cite{coral_benchmark},
Rodinia benchmark suite~\cite{Che:IISWC09}, 
SPEC CPU2000~\cite{henning2000spec},
and two scientific applications (Hercules for earthquake simulation~\cite{aktulga2012pc} and PuReMD for reactive molecular dynamics simulation~\cite{taborda2011cse}). 
We carefully choose 25 applications from the above benchmark suites and scientific applications for testing. The 25 applications are shown in Table~\ref{tab:big_apps_accy}.
We call the 25 applications \textit{big benchmarks} in the rest of the paper. 

To train \name, we use 100 common computation kernels from HackerRank~\cite{hackerrank}.
These kernels are relatively shorter than the big benchmarks,
but these kernels all have explicit verification phases. 

\textbf{Trace Generation.}
We use LLVM-Tracer~\cite{ispass13:shao}, a tool to generate dynamic LLVM IR traces based on LLVM instrumentation.  The trace includes LLVM IR instructions and their operands.


To introduce the execution order information, we define ``chunk"; each chunk is the dynamic instruction trace of a loop or code between two neighbor loops. 
We extend LLVM-tracer to generate a subtrace for each chunk.

\textbf{Regression Model Selection.}
We use 10-fold cross validation to evaluate 18 regression models on the training dataset to select the best models.
These 18 models are Kneighbors Regression, Gradient Boosting Regression, Random Forest Regression, SV Regression, NuSVR Regression, Decision Tree Regression,SGD Regression, Lasso Regression, Elastic Net Regression, Huber Regression, Bayesian Ridge Regression, Passive-Aggressive Regression, Ridge Regression, KernelRidge Regression, TheilSen Regression, RANSAC Regression, Least Square Linear Regression, and MLP Regression. 
We use scikit-learn~\cite{scikit-learn} to implement the models. 
Table~\ref{tab:err_vari} 
shows the top five regression models with the highest prediction accuracy.
The remaining 13 regression models 
are not listed because of their low prediction accuracy. 
\textbf{Tuning Hyperparameters.} Each regression model has multiple hyperparameters.
We leverage ``grid-search''~\cite{bergstra2012jmlr} to decide the values of hyperparameters for training. 

\section{Evaluation}
\label{sec:eval}
We use the trained regression models
to predict the rate of success and 
interruption. 
The SDC rate is simply the result of subtracting the rates 
of success and interruption from one (``1'').
We do not use the models to directly predict the SDC rate, 
because the observed SDC rate ($O_{rate}$ in Equation~\ref{eq:3}) for some applications can be very small (close to 0), which easily makes $|P_{rate} - O_{rate}|/O_{rate}$ in  Equation~\ref{eq:3} larger than 1. As a result, $P_{accuracy}$ is negative, which is counter-intuitive (it should be always non-negative).

We evaluate our models and modeling methods from two perspectives: (1) the modeling accuracy; (2) the contributions of various modeling techniques and model optimization techniques to the modeling accuracy.

\begin{table*}[tbp]
\centering
\scriptsize
\caption{The average prediction accuracy for the three rates (i.e., Success rate=SR, SDC rate=SDCR, and interruption rate=IR). Numbers in the parenthesis are for the variance of the prediction accuracy. Notation: APA=average prediction accuracy, SCK=small computation kernels, HPCB=HPC benchmarks.
}
\label{tab:err_vari}
\begin{tabular}{|p{3cm}|p{2cm}|p{2.1cm}|p{2cm}|p{2cm}|}
\hline
Regression models    & APA for SR on SCK & APA for SR on HPCB & APA for IR on SCK & APA for IR on HPCB  \\
\hline   \hline
SV Regression                & 0.75 (0.15)                                   & 0.72 (0.17)    & 0.71 (0.17)   & 0.67 (0.24)   \\ 
\hline 
Gradient Boosting Regression & 0.81 (0.13)                                  & 0.82 (0.02)   & 0.75 (0.15)   & 0.77 (0.05)                    \\
\hline
Random Forest Regression     & 0.77 (0.14)                                   & 0.74 (0.02)    & 0.72 (0.14)   & 0.71 (0.18)                       \\
\hline
Kneighbors Regression        & 0.74 (0.16)                                   & 0.7 (0.04)  & 0.63 (0.21)   & 0.56 (0.32)                       \\
\hline
NuSVR Regression             & 0.75 (0.21)                                   & 0.74 (0.08)    & 0.74 (0.17)   & 0.66 (0.26)             
\\
\hline
\end{tabular}
\vspace{-10pt}
\end{table*}

\begin{table}[t]
\scriptsize
\captionsetup{justification=raggedright,singlelinecheck=false}
\caption{Feature voting scores for each dimension of the feature vector $\mathcal{F}^{ave}_{30}$.}
\vspace{-5pt}
\label{tab:feature_voting}
\begin{subtable}[h]{0.5\textwidth}
\caption{Feature voting scores for predicting the success rate.}
\begin{tabular}
{|p{3.73cm}p{0.05cm}p{0.05cm}p{0.05cm}p{0.05cm}p{0.05cm}p{0.05cm}p{0.05cm}p{0.05cm}p{0.05cm}p{0.05cm}|} \hline 
Dimension Number                 & 4  & 24 & 8  & 28 & 17 & 12 & 14 & 22 & 18 & 27 \\
Sorted voting score (Smaller is better) & 20 & 22 & 23 & 24 & 25 & 27 & 29 & 29 & 31 & 32 \\
\hline \hline
Dimension Number & 2  & 3  & 23 & 7  & 20 & 16 & 6  & 21 & 13 & 26  \\
Sorted voting score (Smaller is better) & 33 & 39 & 39 & 40 & 43 & 45 & 46 & 48 & 50 & 50 \\
\hline \hline
Dimension Number & 11 & 1  & 30 & 15 & 5  & 10 & 25 & 29 & 9  & 19   \\
Sorted voting score (Smaller is better) & 53 & 54 & 62 & 69 & 70 & 71 & 74 & 74 & 86 & 87  \\
\hline
\end{tabular}
\end{subtable}
\begin{subtable}[h]{0.5\textwidth}
\caption{Feature voting scores for predicting the interruption rate.}
\begin{tabular}{|p{3.73cm}p{0.05cm}p{0.05cm}p{0.05cm}p{0.05cm}p{0.05cm}p{0.05cm}p{0.05cm}p{0.05cm}p{0.05cm}p{0.05cm}|} \hline 
Dimension Number                 & 14 & 18 & 4  & 8  & 27 & 24 & 28 & 7  & 30 & 16 \\
Sorted voting score (Smaller is better) & 20 & 23 & 24 & 27 & 27 & 32 & 32 & 34 & 37 & 38 \\
\hline \hline
Dimension Number & 6  & 17 & 10 & 26 & 12 & 13 & 1  & 3  & 11 & 2   \\
Sorted voting score (Smaller is better) & 39 & 40 & 42 & 43 & 46 & 46 & 47 & 47 & 52 & 53 \\
\hline \hline
Dimension Number & 21 & 19 & 20 & 23 & 5  & 15 & 22 & 25 & 9  & 29   \\
Sorted voting score (Smaller is better) & 53 & 55 & 55 & 56 & 62 & 63 & 69 & 69 & 77 & 87  \\
\hline
\end{tabular}
\end{subtable}
\vspace{-10pt}
\end{table}

\subsection{Prediction Accuracy}
We show the prediction accuracy in Table~\ref{tab:err_vari} for the top five regression models. We have applied feature selection and model tuning techniques to improve the prediction accuracy of these models.
The second and fourth columns of Table~\ref{tab:err_vari}
show the results from cross-validation. 
We use the 100 small computation kernels for training.
The third and fifth columns of Table~\ref{tab:err_vari}
show the results from testing. We use the
big benchmarks for testing. 

\begin{table*}
\centering 
\tiny
\caption{The detailed prediction results for 25 big benchmarks. Notation: SR=Success Rate; SDCR=SDC Rate; IR=Interruption Rate; Pred.=Prediction; Obs.=Observed; Accy=Accuracy.}
\label{tab:big_apps_accy}
\begin{tabular}{|p{1cm}|p{1cm}|p{1cm}|p{1cm}|p{1cm}|p{1cm}|p{1cm}|p{1cm}|p{1cm}|p{1cm}|p{1cm}|p{1cm}|p{1cm}|}
\hline
Big benchmarks & Suite & Program input & Obs. SR & Pred. SR   & Pred. Accy for SR (higher is better)     & Obs. SDCR         & Pred. SDCR  & Pred. Accy for SDCR (higher is better)  &  \textbf{Pred. Accy for SDCR by \bfseries{Trident}} & Obs. IR    & Pred. IR   & Pred. Accy for IR (higher is better)  
\\ \hline \hline
IS          & NAS & Class S & 0.653        & 0.701   & 0.926     & 0.083    & 0.103      & 0.020     &        N/A              & 0.264             & 0.195        & 0.739     \\ \hline
LU          & NAS & Class S & 0.575        & 0.698   & 0.787     & 0.174    & 0.150      & 0.863     &         N/A              & 0.251             & 0.152        & 0.606     \\ \hline
Nn         & Rodinia & filelist_4 5 30 90  & 0.513        & 0.847   & 0.348     & 0.173    & 0.000      & 0.000     &     N/A                  & 0.314             & 0.323        & 0.973     \\ \hline
Myocyte   & Rodinia &  100 1 0 4  & 0.741        & 0.707   & 0.954     & 0.022    & 0.023      & 0.939     &        N/A               & 0.237             & 0.270        & 0.862     \\ \hline
Backprop   & Rodinia &  35536  & 0.670        & 0.528   & 0.788     & 0.016    & 0.111      & -4.946    &       N/A                & 0.314             & 0.361        & 0.850     \\ \hline
CG        & NAS	 &  Class S    & 0.739        & 0.704   & 0.953     & 0.139    & 0.147      & 0.941     &       N/A                & 0.122             & 0.149        & 0.782     \\ \hline
MG           & NAS	 &  Class S & 0.781        & 0.640   & 0.820     & 0.008    & 0.009      & 0.835     &      N/A                 & 0.211             & 0.351        & 0.339     \\ \hline
BT          & NAS	 &  Class S & 0.656        & 0.606   & 0.924     & 0.164    & 0.252      & 0.465     &       N/A                & 0.180             & 0.142        & 0.791     \\ \hline
SP        & NAS	 &  Class S   & 0.385        & 0.375   & 0.974     & 0.306    & 0.255      & 0.832     &      N/A                 & 0.309             & 0.371        & 0.801     \\ \hline
DC        & NAS	 &  Class S   & 0.578        & 0.690   & 0.806     & 0.060    & 0.000      & 0.000     &       N/A                & 0.362             & 0.396        & 0.906     \\ \hline
Lud       & Rodinia	 &  512.dat  & 0.760        & 0.669   & 0.881     & 0.142    & 0.135      & 0.953     &       N/A                & 0.098             & 0.195        & 0.007     \\ \hline
Kmeans     &  Rodinia   &   100  & 0.843        & 0.712   & 0.844     & 0.045    & 0.093      & -0.070    &      N/A                 & 0.112             & 0.195        & 0.256     \\ \hline
sAMG  & CORAL & aniso      & 0.467        & 0.704   & 0.492     & 0.370    & 0.144      & 0.388     &          N/A             & 0.163             & 0.152        & 0.933     \\ \hline
STREAM    &  PARSEC  &	10 20 64 512 512 100 none oput.txt 4   & 0.723        & 0.611   & 0.845     & 0.066    & 0.194      & -0.938    &           N/A            & 0.211             & 0.195        & 0.926     \\ \hline \hline
Libquantum    &  SPEC  &	33 5  & 0.863        & 0.922   & 0.931     & 0.034    & 0.000      & 0.000     & 0.924                & 0.103             & 0.125        & 0.784     \\ \hline
Blackscholes  &  PARSEC  &	in_4.txt  & 0.663        & 0.571   & 0.862     & 0.122    & 0.203      & 0.338     & 0.878                & 0.215             & 0.226        & 0.949     \\ \hline
Sad     &  Parboil	& reference.bin frame.bin      & 0.475        & 0.498   & 0.951     & 0.216    & 0.314      & 0.546     & 0.650                & 0.309             & 0.188        & 0.607     \\ \hline
Bfs-parboil  &  Parboil &	graph_input.dat& 0.496        & 0.686   & 0.617     & 0.131    & 0.010      & 0.079     & 0.967                & 0.373             & 0.304        & 0.815     \\ \hline
Hercules  & CMU  &	scan simple_case.e   & 0.580        & 0.610   & 0.949     & 0.182    & 0.172      & 0.945     & -0.282               & 0.238             & 0.218        & 0.917     \\ \hline
PuReMD    & Purdue Univ.  &	geo ffield control   & 0.350        & 0.492   & 0.594     & 0.090    & 0.021      & 0.232     & 0.610                & 0.560             & 0.487        & 0.870     \\ \hline
Lulesh     & CORAL & -s 1 -p  & 0.634        & 0.444   & 0.701     & 0.120    & 0.258      & -0.148    & -36.400              & 0.246             & 0.298        & 0.788     \\ \hline
Hotspot    &  Rodinia  & 64 64 1 1 temp_64 power_64   & 0.714        & 0.699   & 0.979     & 0.121    & 0.116      & 0.956     & 0.790                & 0.165             & 0.185        & 0.877     \\ \hline
Bfs-rodinia      &  Rodinia  &	graph4096.txt    & 0.655        & 0.700   & 0.932     & 0.124    & 0.048      & 0.389     & 0.792                & 0.221             & 0.252        & 0.859     \\ \hline
Nw     &  Rodinia  &  2048 10 1      & 0.664        & 0.619   & 0.933     & 0.140    & 0.185      & 0.677     & 0.410                & 0.196             & 0.195        & 0.996     \\ \hline
Pathfinder  &  Rodinia   &   1000 10  & 0.623        & 0.797   & 0.721     & 0.231    & 0.055      & 0.236     & 0.687                & 0.146             & 0.149        & 0.983     \\ \hline \hline
Average(var)   &  N/A   &  N/A  & 0.632        & 0.649   & \textbf{0.820(0.02)}     & 0.131    & 0.120      & \textbf{0.211}     & \textbf{-2.725}               & 0.237             & 0.243        & \textbf{0.769(0.05)}   \\
\hline
\end{tabular}
\vspace{-8pt}
\end{table*}

Table~\ref{tab:err_vari} 
shows that among the five regression models, the Gradient Boosting Regression achieves the best prediction accuracy for both small computation kernels and big benchmarks. The prediction accuracy for them is 82\% and 77\%, respectively. 
The variance of prediction accuracy for the Gradient Boosting Regression is smaller than most of other regression models. Hence, the Gradient Boosting Regression is the best. 
For the following experiments, if indicated otherwise, we only show the results of using the Gradient Boosting Regression.

We present more details of the prediction result in Table~\ref{tab:big_apps_accy}. 
The table shows that the prediction accuracy for predicting the success rate is 82\% (on average) with a variation of 0.02;
The prediction accuracy for predicting the interruption rate
is 77\% with a variation of 0.05. 

\textbf{Comparison with the State-of-the-Art.}
We compare our prediction accuracy with that of Trident~\cite{dsn18:li}, a very recent work that uses
analytical models to estimate the SDC rate.
We use the 11 benchmarks 
evaluated in \texttt{Trident}.
We use the same input for the
11 benchmarks as \texttt{Trident} uses.
Table~\ref{tab:big_apps_accy} shows the prediction
accuracy of Trident in the last 12 rows.

Table~\ref{tab:big_apps_accy} shows that the average prediction accuracy of \name for the 11 benchmarks is 38.6\%. 
However, the average prediction accuracy of \texttt{Trident} 
is -272.5\%. 
\name achieves much better prediction accuracy than \texttt{Trident}.

The reason why \texttt{Trident} has relatively low prediction accuracy is as follows. \texttt{Trident}
uses analytical models to reason the possibility of SDC.
To avoid the complexity of reasoning, they do not analyze
all instructions, which results in low prediction accuracy.

\begin{figure}
	\begin{center}
\includegraphics[height=0.165\textheight]{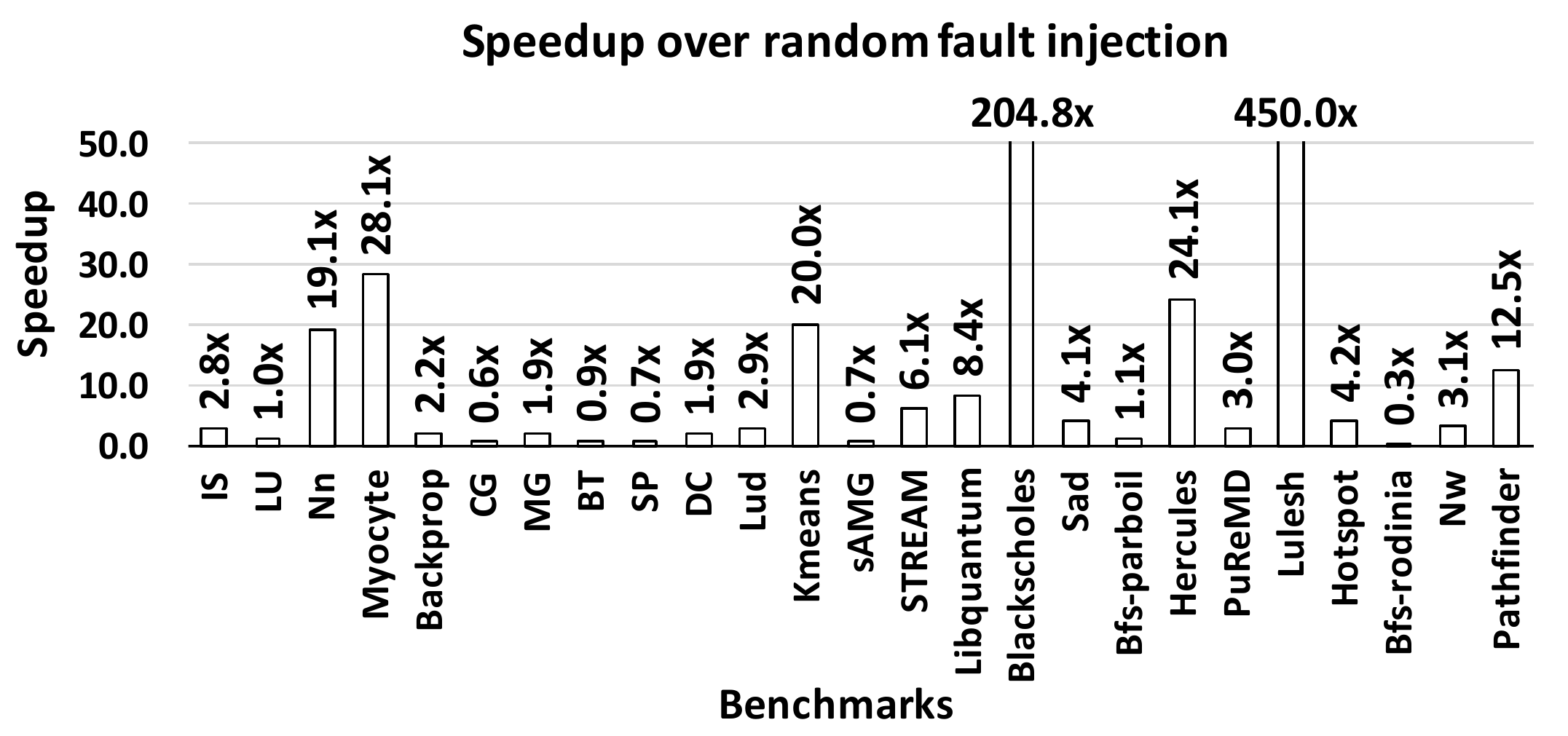} 
		\caption{
        The speedup of using \name over random FI to predict the rate of manifestations. 
        }
		\label{fig:speedup}
	\end{center}
   \vspace{-15pt}
\end{figure}

\subsection{Efficiency Study--Comparing with Random Fault Injection}

We compare the time of using FI and using \name
to predict the rate of manifestations for 
the 25 big benchmarks.
The number of FIs is determined by using 
a statistical approach~\cite{leveugle2009}
with the confidence level of 99\% and
the margin of error 1\%. In particular, 
we use 3000 FIs for each 
benchmark. 
When measuring the time of using \name, 
we measure the time spent on the whole workflow,
including dynamic instruction trace
generation, feature extraction, and making prediction with the trained machine learning model. 

Figure~\ref{fig:speedup} shows the results.
In general, the speedup of using \name over random FI
is up to 450x (see LULESH). Among the 25 benchmarks, \name is faster than random FI for 20 benchmarks.
For one benchmark (LU), \name uses almost the same time as FI.
For the four benchmarks (CG, BT, SP, and bfs), \name is slower, 
because of the time-consuming trace generation. 
We hope to improve the performance of the trace generation by using trace compression in the future.

\subsection{Feature Selection and Analysis}
We use the feature selection technique (i.e., the voting strategy) discussed in Section~\ref{sec:featuer_selection}
to select features.
We analyze the feature selection result in this section. 

Table~\ref{tab:feature_voting} shows the global indexes for all features.  
Table~\ref{tab:feature_voting}.a
reveals that the 4th dimension (the memory-related instructions), 24th dimension (the memory-related instructions in bigram), and 8th dimension (the pattern of overwriting) in $\mathcal{F}^{ave}_{30}$ rank the highest; 
Table~\ref{tab:feature_voting}.b reveals that 
the 14th dimension (the memory-related instructions in bigram),
18th dimension (the pattern of overwriting in bigram),
and 4th dimension (the memory-related instructions)
in $\mathcal{F}^{ave}_{30}$ rank the highest.
Those dimensions are the memory-related instructions, which seem to matter most to the application resilience.

In addition, both tables reveal that 
the 9th dimension (i.e., the pattern of dead location), 19th dimension (i.e., the pattern of dead location in bigram), and 29th 
dimension (i.e., the pattern of dead location in bigram) rank relatively low. 
This result indicates that the feature of dead location seems to contribute less to application resilience than the other features.

\subsection{Evaluation of Model Tuning and Feature Construction Optimization}

\begin{figure}
	\begin{center}
\includegraphics[height=0.21\textheight]{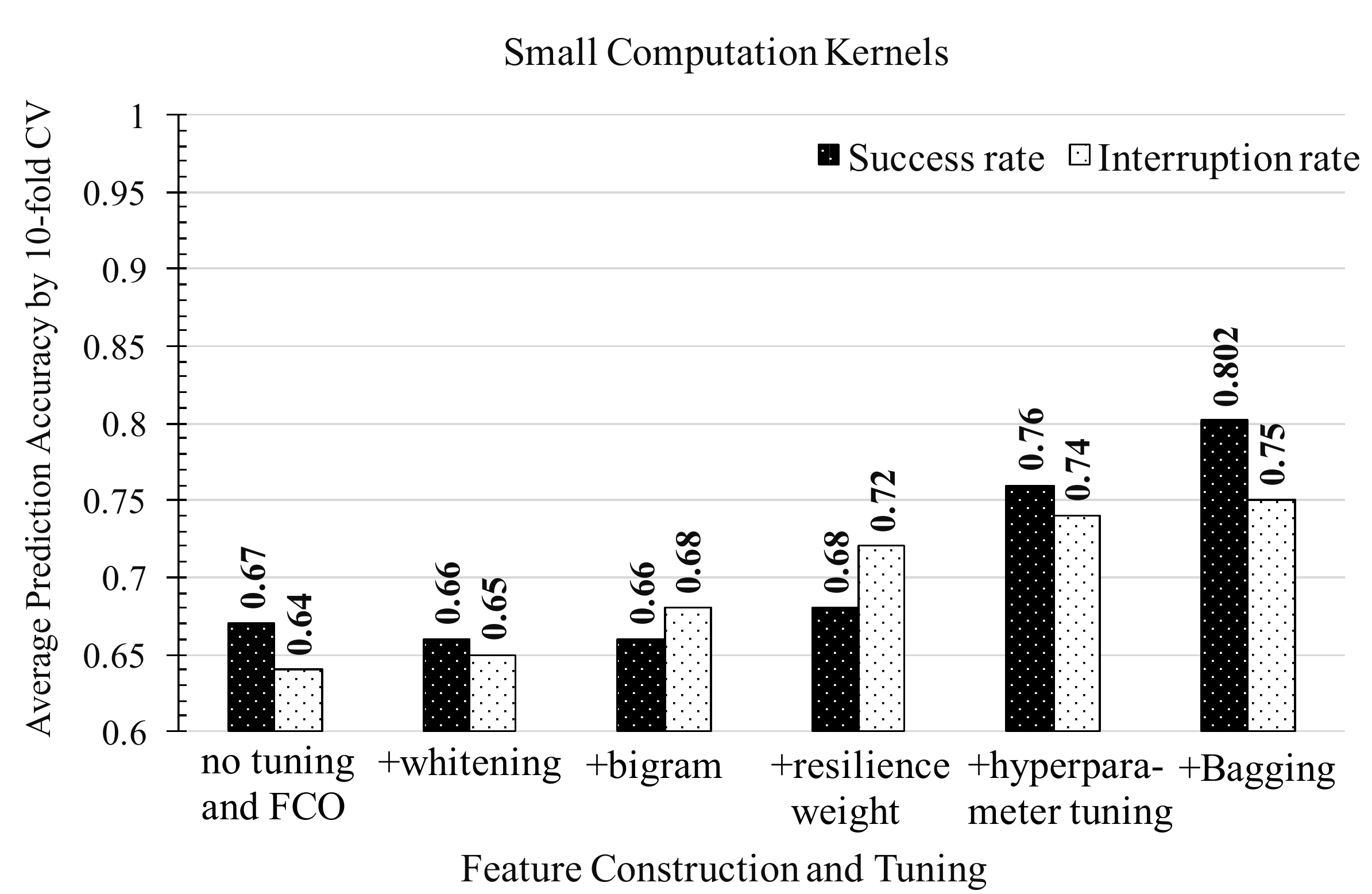} 
		\caption{Evaluating the impact of model tuning and feature construction optimization on the prediction accuracy for the three rates. Notation: FCO = ``feature construction optimization''.
        }
		\label{fig:tuning_process}
	\end{center}
    \vspace{-15pt}
\end{figure}

We study the impact of our model tuning (whitening, bagging and tuning hyperparameters) and feature construction techniques (bigram and resilience weight) on the model accuracy. We use the Gradient Boosting Regression model and 100 small computation kernels (for training) for our study. We start with the model without any of the five techniques, and then apply them one by one. 

Figure~\ref{fig:tuning_process} shows the results. We can see that the prediction accuracy keeps increasing after we apply those techniques one by one. This demonstrates the effectiveness of our techniques. 
Among the five techniques, the most effective ones are resilience weight, hyperparameters tuning, and bagging when predicting the success rate, and bigram and resilience weight when predicting the interruption rate. 

We notice that introducing bigram,
the average prediction accuracy is not increased when predicting the success rate. However, examining individual computation kernels, we find that the prediction accuracy for 71\% of kernels becomes better, with up to 26\% improvement in the prediction accuracy. There are two outliers that largely decrease prediction accuracy after applying bigram.
Furthermore, when predicting the interruption rate, the average prediction accuracy increases 3\% after applying bigram into features. 3\% is a large improvement in the machine learning field.
Hence, we conclude that using bigram is very helpful to improve the modeling accuracy.

\section{Related Work}
\textbf{Using Machine Learning to Address Resilience Problems.}
There are a couple of recent efforts that use machine learning 
~\cite{pact15:mitra,cgo16:laguna,vishnu2016ipdps,das2018hpdc,sc15:ashraf,nie2018dsn} to address resilience problems.
Mitra et al.~\cite{pact15:mitra} build a regression model to predict anomaly output of an application, given a certain combination of input parameters to the application. Laguna et al.~\cite{cgo16:laguna} train a machine learning classifier IPAS. IPAS learns which instructions can have a high likelihood of leading to a silent output corruption.  IPAS duplicates those instructions to mitigate the effect of silent output corruption. 
Vishnu et al.~\cite{vishnu2016ipdps}
use attributes including system and
application states to predict whether
a multi-bit error will lead to corrupted output. 
Desh~\cite{das2018hpdc} predict 
node failures by training a recurrent neural network model
using system logs.
Nie et al.~\cite{nie2018dsn} use system characteristics such as temperature, power consumption, application states as features to 
predict the occurrence of GPU errors.
\name is the first work applying machine
learning to predict the rate of manifestations. 




\textbf{Random FI.}
This is the most common method is to study application resilience 
~\cite{sc14:cher,europar14:calhoun,ispass17:hari,bifit:SC12,hpca17:de,dsn14:luo,micro94:karlsson,dsn24:gemfi,asplos08:Manlap,ispass14:bo,toc95:kanawati}.
Typically application-level FI has to be performed many times to 
ensure statistical significance. Some research prunes unnecessary FI to reduce FI efforts.
Hari et al. ~\cite{asplos12:hari} 
explore instruction equivalence for selective FI.
They further reduce FI positions by leveraging the equivalence of intermediate states in execution and instruction-level approximate computing ~\cite{isca14:hari,micro16:hari}. 
Our work tries to address the inefficiency of FI to study application resilience. But the above existing work is complementary to our work. 

\textbf{Error Propagation Analysis.}
Application level error propagation has been widely studied.
Li et al.~\cite{sc16:guanpeng} implement a FI tool to study error propagation in GPU applications, and  
\texttt{Trident}~\cite{dsn18:li}, a three-level error propagation model to predict SDC probabilities of programs.
Calhoun et al.~\cite{hpdc17:Calhoun} study how corruption state changes due to error propagation at the instruction and application variable level for three applications. Ashraf et al.~\cite{sc15:ashraf} propose an error propagation model to study error propagation for MPI applications. 
Our work does not focus on error propagation, but includes an N-gram based technique to embed the execution order information into the feature vector to consider the effects of error propagation.
\section{Conclusions}
As supercomputers increase in size and complexity,
the rate of transient faults is expected to 
increase and becomes a severe problem threatening computation correctness. 
Techniques to understand the 
manifestation of transient faults
become increasingly important to ensure result correctness for those applications running on supercomputers. 
This paper introduces \name, a machine learning based approach to predict the rate of manifestations of transient faults. 
We train \name on 100 small computation kernels
and test on 25 big benchmarks using 
features highly related to application resilience.
We test 18 regression models and find the Gradient Boosting Regression the best machine learning model for predicting 
the rate of manifestations of transient faults in terms of prediction accuracy.

\bibliographystyle{ACM-Reference-Format}
\bibliography{paper} 

\end{document}